\documentclass[copyright,creativecommons]{eptcs}
\renewcommand{\event}{S. Barry Cooper \& Vincent Danos (eds.):\\
Fifth Workshop on Developments in Computational Models\\
---Computational Models From Nature}
\providecommand{\volume}{??}
\providecommand{\anno}{20??}
\providecommand{\firstpage}{1}
\usepackage{amssymb}

\title{Circular languages generated
by complete splicing systems and pure unitary languages
\thanks{Partially supported by
the {\bf MIUR} Project
{\it ``Mathematical aspects and emerging
applications of automata and formal languages''}
(2007), by the {\bf ESF} Project
{\it ``Automata: from Mathematics to Applications
(AutoMathA)''} (2005-2010),
by the $60 \%$ Project
{\it ``Propriet\`a strutturali e nuovi modelli di
rappresentazione nella teoria dei linguaggi formali''}
(University of Salerno, 2007) and
by the $60 \%$ Project
{\it ``Estensioni della teoria dei linguaggi formali
e loro propriet\`a strutturali''}
(University of Salerno, 2008).}}
\author{Paola Bonizzoni
\institute{Dipartimento di Informatica Sistemistica e Comunicazione \\
Universit\`a degli Studi di Milano - Bicocca \\
Viale Sarca 336,
20126 Milano - Italy }
\email{bonizzoni@disco.unimib.it}
\and
Clelia De Felice, Rosalba Zizza
\institute{Dipartimento di Informatica ed
Applicazioni, \\
Universit\`a di Salerno, via Ponte Don Melillo, \\
84084 Fisciano (SA), Italy}
\email{\{defelice,zizza\}@dia.unisa.it}
}

\newtheorem{propo}{Proposition}[section]
\newtheorem{de}{Definition}[section]
\newtheorem{teo}{Theorem}[section]
\newtheorem{lemm}{Lemma}[section]
\newtheorem{rema}{Remark}[section]
\newtheorem{exa}{Example}[section]
\newtheorem{coro}{Corollary}[section]
\def\petitcarre{\vrule height4pt width 4pt depth0pt}
\def\enddim{\relax\ifmmode\eqno{\hbox{\petitcarre}}
\else
{\unskip\nobreak\hfil\penalty50
   \hskip2em\hbox{}\nobreak\hfil
   \petitcarre
   \parfillskip=0pt \finalhyphendemerits=0
  \par\medskip}\fi}

\newcommand{\1}{\hspace{0 mm}^\sim \hspace{-0.2 mm}}

\def\titlerunning{Circular languages generated
by complete splicing systems and pure unitary languages}
\def\authorrunning{Paola Bonizzoni \& Clelia De Felice \& Rosalba Zizza}

\begin{document}
\maketitle

\begin{abstract}

{\it Circular splicing systems} are a formal model of a
generative mechanism of circular words, inspired by a
recombinant behaviour of circular DNA. Some unanswered
questions are related to the computational power of such systems,
and finding a characterization of the class of
circular languages generated by circular splicing systems
is still an open problem. In this paper we
solve this problem for {\it complete systems}, which are
special {\it finite circular splicing
systems}. We show that a circular
language $L$ is generated by a complete system if and
only if the set $Lin(L)$ of all words corresponding
to $L$ is a {\it pure unitary language} generated by
a set closed under the conjugacy
relation. The class of pure unitary languages was
introduced by A. Ehrenfeucht, D. Haussler, G. Rozenberg
in 1983, as a subclass of the class of context-free languages,
together with a characterization of regular pure unitary languages
by means of a decidable property.
As a direct consequence, we characterize (regular)
circular languages generated by complete systems.
We can also decide whether the language generated
by a complete system is regular.
Finally, we point out that complete systems
have the same computational power
as {\it finite simple systems},
an easy type of circular splicing system
defined in the literature from the very beginning,
when only one rule is allowed.
From our results on complete systems, it follows that
finite simple systems generate a class of context-free
languages containing non-regular languages, showing the incorrectness
of a longstanding result on simple systems.
\end{abstract}
%----------------------

\section{Introduction} \label{introduction}

The notion of {\it linear splicing systems} was first
introduced in \cite{h87}, where Head modelled a
recombinant behaviour of DNA molecules
as a particular operation between words in a formal language.
Subsequently, {\it circular splicing systems}
were introduced in \cite{H92}
along with various open problems related to their computational
power. In the circular context, the splicing operation
acts on two circular DNA molecules by means of
a pair of restriction enzymes as follows.
Each of these two
enzymes is able to recognize a pattern inside one
of these two circular DNA molecules and to cut the
molecule in the middle of such a pattern. Two linear
molecules are produced and then they are pasted together
by the action of ligase enzymes. Thus a new circular DNA
sequence is generated \cite{h87,hb,book}.
For instance, circular splicing models the integration
of a plasmid into the DNA of a host bacteria.
A language-theoretic operation
can be defined and, depending on whether or not these
ligase enzymes substitute the recognized
pattern, we have Pixton's definition or Head's and P\u{a}un's
definition.

In the circular context we deal with {\it circular words},
i.e., equivalence classes with respect to the conjugacy
relation
$\sim$, defined by $xy \sim yx$, for $x,y \in A^*$
\cite{lot}.
Loosely speaking, a circular word is a word $w$
written on a circle
and all the words that are equivalent to $w$
can be obtained by reading the letters on the circle,
starting from any point on it.
We can then consider sets of circular words or
{\it circular languages}.
In short, a circular splicing system is a triple $(A,I,R)$ where
$A$ is a finite alphabet, $I$ is the {\it initial} circular
language and $R$ is the set of {\it rules},
having the form
$u_1 \# u_2 \$ u_3 \# u_4$, with $u_i \in A^*$,
$1 \leq i \leq 4$.
Splicing rules are
iteratively applied starting from $I$. The corresponding
{\it circular splicing language} is the smallest
language which contains $I$ and is invariant under iterated
splicing by rules in $R$.
Finding a characterization of the class of
circular languages generated by circular splicing systems
is still an open problem. Partial results are known
for Pixton (resp. P\u{a}un) circular
splicing systems $S = (A,I,R)$ with $I$ and $R$
satisfying additional assumptions, namely $R$ is
assumed to be a {\it reflexive},
{\it symmetric} set of rules and {\it self-splicing}
is allowed \cite{hb,pixDAM}.

In this paper we consider classes of splicing systems
where the splicing operation is of a simpler
form and for which finding a relationship with regular
circular languages is a quite natural
question investigated in this research area.
Precisely, we consider P\u{a}un systems
$S$ with both $I$ and $R$ finite sets
({\it finite} P\u{a}un systems)
and such that $a \# 1 \$ b \# 1 \in R$,
for each $a,b \in A$,
where $1$ is the empty word and $A$ is the alphabet
of $I$.
In this case the splicing operation is very easy:
it applies to any pair of circular words, $\1 xa, \1 yb$,
and gives as a result the circular word $\1 xayb$.
Therefore, the splicing operation
can be seen as a concatenation followed by a closure under 
the conjugacy relation.
We have called these systems {\it complete}.

The main contribution of this paper is the characterization
of the computational power of complete systems.
Indeed, we show that
a circular language $L$ is generated
by a complete system if and only if $Lin(L)$, the set of
all words such that the corresponding circular words are
in $L$, is a {\it pure unitary language}
generated by a set closed under the conjugacy relation.

The class of pure unitary languages is a subclass
of the class of context-free languages, introduced
in \cite{EHRtcs}. Equivalently, a language is pure unitary
if it is obtained by applying the {\em iterated insertion} operation,
starting with a finite set $Y$ \cite{Hau}.
Thus we show that circular splicing and iterated insertion
are closely related.
In the same paper \cite{EHRtcs}, the authors
also characterized regular pure unitary languages by means
of a decidable property. As a consequence, we characterize
regular circular languages generated by complete systems
and we can also decide whether a complete system
generates a regular circular language.

We also show that complete systems have the
same computational power as
circular {\it simple systems}
with only one rule, i.e., finite P\u{a}un
systems with $R=\{a \# 1 \$ a \# 1\}$, $a \in A$.
As a consequence, we characterize (regular) circular
languages generated by circular
simple systems with only one rule.
The special class of simple
systems was first considered
in \cite{ssd} using Head's definition and then in
\cite{CS2} by taking into account P\u{a}un's definition,
as the circular counterpart of the case of the linear
splicing investigated in \cite{MPRS}.
In \cite{ssd}, the authors claimed that Head
simple systems always generate regular circular languages.
In this paper we show that there are simple systems
generating non-regular context-free circular languages
and we give new contributions towards the solution of
the still open problem of finding a characterization of the
class of circular languages generated by finite circular simple
splicing systems.
(As a matter of fact, an example of a non-regular
circular language generated by a simple system has been
also reported in \cite{survey}, see Example \ref{ultimo}).

A still unanswered question is whether
the above-mentioned characterization of regularity can be generalized
to larger classes of circular splicing languages, in particular
to the class of simple systems.

The paper is organized as follows. In Section \ref{basic}
we gathered basics on (circular) words and languages,
circular splicing, CSSH-systems and simple systems. We also
conclude the section with a short description
of how state of the art the open problems are.
In Section \ref{PUL} we state
some known and new results on pure unitary languages.
Section \ref{MR} is devoted to our
main results.
All the proofs have been omitted for space constraint but they can
be found in an extended version of this paper \cite{estesa}.
The main results of this paper were also
communicated at {\it AutoMathA 2009} (Liege, Belgium,
8-12 June 2009).

%---------------------------------------------------------
\section{Basics} \label{basic}

\subsection{(Circular) words and languages} \label{circularw}

We denote by $A^*$ the free monoid over a finite alphabet $A$
and we set $A^+=A^* \setminus 1$, where $1$ is the empty
word. For a word $w \in A^*$, $|w|$ is the length of $w$ and
$|w|_a$ is the number of occurrences of $a$ in $w$, $a \in A$.
We also set $alph(w) = \{a \in A ~ | ~ |w|_a > 0 \}$.
A word $x \in A^*$ is a {\it factor} of $w \in A^*$ if
$u_1,u_2 \in A^*$ exist such that $w=u_1xu_2$.
For a subset $Y$ of $A^*$, $|Y|$ is the cardinality of $Y$
and $alph(Y) = \cup_{y \in Y} alph(y)$.
We denote by $Fin$ (resp. $Reg$) the class of finite
(resp. regular) languages over $A$.
Given $w \in A^*$, a circular word
$\1 w$ is the equivalence class of $w$ with respect to
the {\em conjugacy} relation $\sim$ defined by $xy \sim yx$,
for $x,y \in A^*$ \cite{lot}.
The notations $|\1 w|$, $|\1 w|_a$,
$alph(\1 w)$ will be defined
as $|w|$, $|w|_a$, $alph(w)$,
for any representative $w$ of $\1 w$.
We will often use the notation $w$ for a circular word
$\1 w$.
Let $\1 A^*$ denote the set of all circular words
over $A$, i.e., the quotient of $A^*$ with respect to $\sim$.
Given $L \subseteq A^*$, $\1 L=\{ \1 w ~|~ w \in L\}$
is the {\it circularization} of $L$ whereas, given
a {\it circular language} $C \subseteq \1 A^*$,
every $L \subseteq A^*$ such that
$\1 L=C$ is a {\it linearization} of $C$.
In particular, a linearization of $\1 w$ is a linearization
of $\{\1 w \}$, whereas the {\em full linearization}
$Lin(C)$ of $C$ is defined by $Lin(C)=
\{w \in A^* ~|~ \1 w \in C \}$.
We will often write $\1 w$ instead of $\{\1 w \}$ and $L$
instead of $\1 L$, for a set of letters
$L \subseteq A$. Given a family of languages
$FA$ in the Chomsky hierarchy, $FA^\sim$ is the set of all
those circular languages $C$ which have some linearization
in $FA$. Thus, $Reg^\sim = \{ C
\subseteq \1 A^* ~|~ \exists L \in Reg : ~ \1 L=C \}$.
If $C \in Reg^\sim$ then $C$ is a {\it regular circular language}.
It is classically known that given a regular
(resp. context-free) language $L \subseteq A^*$,
$Lin(\1 L)$ is regular (resp. context-free) \cite{hum}.
As a result, given $C \subseteq \1 A^*$,
we have $C \in Reg^\sim$ (resp. $C$ is a {\it context-free
circular language}) if and only if $Lin(C)$ is
regular (resp. context-free) \cite{hb}.

\subsection{Circular splicing} \label{circulars}

In this paper we deal with the definition of
the circular splicing operation given in \cite{hb}.
The corresponding circular splicing systems are named
here {\em P\u{a}un circular splicing systems} since
they are the counterpart of P\u{a}un linear
splicing systems in the circular context.

\smallskip
{\bf \tt P\u{a}un's definition \cite{hb}.}
A {\em P\u{a}un circular splicing system} is a triple $S = (A,I,R)$,
where $A$ is a finite alphabet, $I$ is the {\em initial} circular
language, with $I \subseteq \1 A^*$ and $R$ is the set
of the {\em rules}, with $R \subseteq A^* \# A^* $$\$ A^* \# A^*$
and $\#, \$ \not \in A$. Then, given a rule
$r=u_1 \#u_2 \$ u_3 \# u_4$ and circular words
$\1 w'$, $\1 w''$, $\1 w$, we set $(\1 w', \1 w''){\vdash}_r \1 w$
if there are linearizations $w'$ of $\1 w'$,
$w''$ of $\1 w''$, $w$ of $\1 w$ such that $w'= u_2xu_1$,
$w''= u_4yu_3$ and $w = u_2xu_1u_4yu_3$.
If $(\1 w', \1 w''){\vdash}_r \1 w$ we say that $\1 w$ is
generated (or spliced) starting with
$\1 w'$, $\1 w''$ and by using a rule $r$. We also say
that $u_1u_2$, $u_3u_4$ are {\em sites} of
splicing.
\medskip

We recall that in the original definition of circular
splicing given in \cite{hb}, rules in $R$
could be used in two different ways.
One way has been described above,
while the other, known as
{\it self-splicing}, will not be considered here.
From now on, ``splicing system'' will be
synonymous with ``circular splicing system''.

Given a splicing system $S$ and a circular
language $C \subseteq \1 A^*$, we set
$\sigma'(C)=\{w \in
\1 A^* ~|~ \exists w', w'' \in C, \exists r \in R : \;
(w',w''){\vdash}_r ~w \}$. We also define
$\sigma^0(C)=C$,
$\sigma^{i+1}(C) = \sigma^i(C) \cup \sigma'(\sigma^i(C))$, $i \geq 0,$
and $\sigma^*(C) = \bigcup_{i \geq 0} \sigma^i(C)$.
Then, $L(S)=\sigma^*(I)$ is
the circular language {\it generated} by $S$.
A circular language $C$ is
{\it P\u{a}un generated} (or $C$ is a {\it circular
splicing language}) if a
splicing system $S$ exists such that $C = L(S)$.
In the sequel $C(Fin, Fin)$ denotes the class of
the circular languages generated by finite P\u{a}un
splicing systems, where
$S = (A,I,R)$ is a {\it finite} splicing system
if $I$ and $R$ are both finite sets.

As observed in \cite{hb}, we may assume that
the set $R$ of the rules in a splicing system
$S = (A,I,R)$ satisfies additional conditions,
having also a biological counterpart.
Namely, we may assume that $R$ is
{\it reflexive} (i.e., for each
$u_1 \# u_2 \$ u_3 \# u_4 \in R$, we have
$u_1 \# u_2 \$ u_1 \# u_2$, $u_3\#u_4 \$ u_3 \#u_4 \in R$)
or $R$ is {\it symmetric} (i.e., for each
$u_1 \# u_2 \$ u_3 \# u_4 \in R$, we have
$u_3 \# u_4 \$ u_1 \#u_2 \in R$).
We do not assume that $R$ is reflexive.
On the contrary, we notice that, in view of
the definition of circular splicing, if
$(w',w''){\vdash}_r ~ w$, with
$r = u_1 \# u_2 \$ u_3 \# u_4$, then
$(w'',w'){\vdash}_{r'} ~ w$, with
$r' = u_3 \# u_4 \$ u_1 \#u_2$. Consequently,
$L(S) = L(S')$, where $S' = (A,I,R')$ and
$R' = R \cup \{u_3 \# u_4 \$ u_1 \#u_2 ~|~
u_1 \# u_2 \$ u_3 \# u_4 \in R \}$. Hence,
in order to find a characterization of the
circular splicing languages, there is no loss of
generality in assuming that $R$ is symmetric.
Thus, in what follows, we assume that
$R$ is symmetric. However,
for simplicity, in the examples of P\u{a}un systems,
only one of either $u_1 \# u_2 \$ u_3 \# u_4$
or $u_3 \# u_4 \$ u_1 \#u_2$ will be reported
in the set of rules.

As already said, in this paper we consider some
special classes of finite circular splicing systems.
In detail, a {\it P\u{a}un circular semi-simple splicing
system} ({\it CSSH system} for short) is a finite splicing
system $S = (A,I,R)$ such that, for each
$u_1 \# u_2 \$ u_3 \# u_4 \in R$, we have
$|u_1u_2| = |u_3u_4| = 1$ \cite{CMS}.
CSSH systems have
been considered in \cite{CMS,CS2,ssd}, once again as
the circular counterpart of linear semi-simple splicing
systems introduced in \cite{GP}.
If $u_1u_2 = u_3u_4 \in A$ then $S$ is
a {\em simple system} \cite{CS2}.
Thus, there are four types of rules, namely
$a_i \# 1 \$ a_j \# 1$, $a_i \# 1 \$ 1 \# a_j$,
$1 \# a_i \$ a_j \# 1$ and $1 \# a_i \$ 1 \# a_j$,
with $a_i, a_j \in A$.
Furthermore, since $R$ is symmetric,
if $a_i \# 1 \$ 1 \# a_j \in R$ then we also have
$1 \# a_j \$ a_i \# 1 \in R$.
So, using the terminology of \cite{CMS,CS2}, a
{\it $(1,3)$-CSSH system}
(resp. {\it $(1,3)$-circular simple system})
is a CSSH system (resp. circular simple system)
where each rule has the form $a_i \# 1 \$ a_j \# 1$,
with $a_i, a_j \in A$. A
{\it $(2,3)$-CSSH system}
(resp. {\it $(2,3)$-circular simple system})
is a CSSH system (resp. circular simple system)
where each rule has the form $1 \# a_i \$ a_j \# 1$,
with $a_i, a_j \in A$. Finally, a
{\it $(2,4)$-CSSH system}
(resp. {\it $(2,4)$-circular simple system})
is a CSSH system (resp. circular simple system)
where each rule has the form $1 \# a_i \$ 1 \# a_j$,
with $a_i, a_j \in A$.
Notice that in a $(1,3)$-CSSH system,
circular splicing can be rephrased as
follows: given a rule $a_i \# 1 \$ a_j \# 1$
and two circular words $\1 xa_i$, $\1 ya_j$,
the circular splicing yields as a result $\1 xa_i ya_j$.
In what follows and in order to abbreviate, we will write $(a_i,a_j)$
to denote a rule $a_i \# 1 \$ a_j \# 1$
in a $(1,3)$-CSSH system. We will focus on a special subclass
of these systems, defined in Section \ref{MR} and named
{\it complete systems}.

\subsection{State of the art}

In this section we will give a brief description of how state
of the art the open problems on circular splicing are.
The main result concerning the computational power of circular
splicing systems states that if $S = (A,I,R)$
is a P\u{a}un circular splicing system such that
$I \in FA^\sim$, where $FA$ is a full abstract
family of languages which is closed under cyclic closure
(i.e., if $L \in FA$ then $Lin(\1 L) \in FA$),
$R$ is a finite, reflexive and symmetric
set of rules and self-splicing is used,
then $L(S) \in FA^\sim$ \cite{hb}.
Other results have been proved for finite circular
splicing systems \cite{dna6,rairo,dam}. In particular, it is known that
$C(Fin,Fin)$ contains regular circular languages
(see \cite{rairo}), context-free circular languages
which are not regular (see \cite{rairo,ssd}), context-sensitive
circular languages which are not context-free (see \cite{Fagnot})
and there exist regular circular languages which are not
in $C(Fin,Fin)$ (see \cite{rairo}).
However, the problem of characterizing
circular languages in $Reg^\sim \cap C(Fin,Fin)$
remains open. We do not even know if, given $L$ in $C(Fin,Fin)$,
it is decidable whether $L$ is regular.
A characterization of $C(Fin,Fin)$ (and of 
$Reg^\sim \cap C(Fin,Fin)$) has been obtained 
for languages over a
one-letter alphabet in \cite{rairo,dam}.

Concerning CSSH systems, in \cite{CS2},
the authors compared the classes of circular languages
generated by $(i,j)$-circular simple systems,
for different values of $(i,j) \in \{(1,3),(2,4), (2,3)\}$,
i.e., for different positions of the letter in the rule.
They proved that the class of circular languages
generated by $(1,3)$-circular simple systems (resp. $(2,3)$-circular
simple systems) is equal
to the class of languages generated by $(2,4)$-circular simple
systems (resp. $(1,4)$-circular simple systems), whereas
the class of circular languages generated by $(1,3)$ and $(2,3)$-circular
simple systems are not comparable.
An analogous viewpoint was adopted for P\u{a}un circular
semi-simple splicing systems in \cite{CMS} where the authors
highlighted further differences between
circular simple and CSSH systems.
Indeed, they proved that the class of circular languages
generated by $(1,3)$-CSSH systems and
the class of circular languages generated by $(2,4)$-CSSH systems
are not comparable. Finally, in \cite{DFFZtcs} the authors proved that
the class of circular languages generated by $(2,4)$-CSSH systems 
is the class of the reversal of the circular languages generated 
by $(1,3)$-CSSH systems. Loosely speaking, the reversal
of a (circular) word is the (circular) word written backwards 
and the reversal of a 
(circular) language $L$ is the (circular) language consisting of 
the reversals of all its (circular) words. 

In \cite{Fagnot}, the author claimed that the class of
circular languages generated by CSSH systems is contained
in the class of context-free circular languages.
Marked systems, i.e., CSSH systems satisfying additional
hypotheses, were introduced in \cite{DFFZ,DFFZtcs}
with a characterization of the corresponding
regular circular languages generated.
This characterization was reviewed in a graph theoretical
setting in \cite{survey}.
However, a still open problem is to find
a characterization of the class of regular circular
languages generated by CSSH systems.

\begin{rema} \label{controesempio}
{\rm As stated in the introduction, the notion of simple splicing
systems has been originally considered using the Head splicing
operation. A Head circular splicing system
$S_H = (A,I,T,P)$ is defined by giving
a finite alphabet $A$, the initial set
$I \subseteq \1 A^*$, the set $T$ of triples,
$T \subseteq A^* \times A^* \times A^*$,
and where $P$ is a binary relation on $T$ such
that, for each $(p,x,q), (u,y,v) \in T$,
$(p,x,q)P(u,y,v)$ if and only if $x=y$.
Head defined circular splicing
as an operation on two circular words
$\1 hpxq$, $\1 kuxv \in \1 A^*$
performed by two triples $(p,x,q)$,$(u,x,v)$
and producing $\1 hpxvkuxq$.
The word $x$ is called a {\em crossing} of the triple.
In \cite{rairo}, the authors proved that there
is a canonical transformation of a Head system
$S_H = (A,I,T,P)$ into a P\u{a}un system $S = (A,I,R)$
such that $S$ and $S_H$ generate the same language.
This transformation is defined by
$R = \{px \# q \$ ux \# v ~|~ (p,x,q), (u,x,v) \in T \}$
(Proposition 3.1 in \cite{rairo}).
We notice that this transformation defines
a bijection between the class
of the circular languages generated by
$(1,3)$-circular simple systems $S = (A, I, R)$, with
$R = \{ (a,a) \}$ and that of
circular languages generated by Head splicing systems
$S_H = (A, I, T, P)$ with
$T = \{ (1,a,1) \}$, $(1,a,1)P(1,a,1)$.
On the other hand, in \cite{ssd}, the authors discussed Head
systems $S_H = (A, I, T, P)$ and the corresponding
splicing operation was named {\it action SA1} of
$S_H$. Furthermore, a triple in $T$ of the form
$(1,a,1)$, $a \in A$, was said to have {\it null
context} and crossing of length one. The first
part of Theorem 3.4 in \cite{ssd} claims the
regularity of the circular splicing language under
action SA1 of a splicing system $S_H = (A, I, T, P)$
with $I$ finite and triples in $T$ having null context
and crossings of length one. On the contrary,
Example \ref{ultimo} shows that
$(1,3)$-circular simple systems $S$ exist
with $L(S)$ being a non-regular language, so disproving the
above-mentioned Theorem 3.4 in \cite{ssd}.}
\end{rema}

%-----------------------------------------------------------

\section{Pure unitary languages} \label{PUL}

In this section, we deal with a class of languages, named
{\it pure unitary languages}, already considered
in \cite{EHRicalp,EHRtcs,Hau} with the aim
of finding conditions under which a context-free grammar
will generate a regular language. Pure unitary languages
(named insertion languages in \cite{EHRicalp}) are a simple class of
``generalized Dyck languages'' and they can be defined
in several ways. Here we follow the viewpoint adopted in \cite{Hau},
where these languages are defined by means of the
operation of {\it iterated insertion}. We report
some of their known properties and we also give a condition
under which a pure unitary language is closed under the conjugacy relation
(Lemma \ref{closure}).
The operations of {\it insertion} and iterated insertion
are variants of classical operations on formal languages.
We recall their definitions below.

\begin{de} \cite{Hau}
Given $Z, Y \subseteq A^*$, the operation of {\it insertion},
denoted by $\leftarrow$, is defined by $Z \leftarrow Y =
\{z_1yz_2 ~|~ z_1z_2 \in Z \mbox{ and } y \in Y \}$. The
operation of {\it iterated insertion}, denoted
by $\leftarrow_*$, is defined inductively from the
operation of insertion by $Y^{\leftarrow_0} = \{1 \}$,
$Y^{\leftarrow_{i+1}} = Y^{\leftarrow_{i}} \leftarrow Y$
and $Y^{\leftarrow_{*}} = \cup_{i \geq 0} Y^{\leftarrow_{i}}$.
\end{de}

\begin{de} \label{unitary} \cite{EHRtcs,Hau}
A language $L$ is a pure unitary language
if $L = Y^{\leftarrow_{*}}$ with $Y$ being a
finite set.
\end{de}

Since $Y^{\leftarrow_{*}} = (Y \setminus 1)^{\leftarrow_{*}}$,
in what follows, we assume $Y \subseteq A^+$.
In \cite{EHRtcs} the authors stated
a characterization of regular pure unitary languages by means
of a decidable property. This result is partially reported
below.

\begin{teo} \label{RVP} \cite{EHRtcs}
Let $Y$ be a finite set such that
$alph(Y) = A$. Then
$L = Y^{\leftarrow_{*}}$ is regular if and only if
$Y$ is {\rm subword unavoidable}
in $A^*$, i.e., there exists
a positive integer $k$ such that any word $u \in A^*$,
with $|u| \geq k$, contains as a factor a word of $Y$.
For any regular set $R \subseteq A^*$,
it is decidable whether or not $R$ is subword unavoidable
in $A^*$.
\end{teo}

The construction of a grammar generating $Y^{\leftarrow_{*}}$
is folklore and is reported below.
As usual, here and from now on we denote
by $L(G)$ the language generated by a context-free
grammar $G$.
Let $w = a_{i_1} \cdots a_{i_h} \in Y$,
where $\{a_{i_1}, \ldots , a_{i_h} \}$ is
a multiset of elements in $A$.
Then, we define the production $p_{w}=
X \rightarrow X a_{i_1} X a_{i_2} \cdots
X a_{i_h} X$. We set $G_{Y} = (\{X \},A,P,X)$,
$P = \{ X \rightarrow 1 \} \cup \{p_{w} ~|~ w \in Y \}$
and we shall call $G_{Y}$ the {\it pure unitary grammar}
associated with $Y$.
Our main result follows from the relation
$Y^{\leftarrow_{*}} = L(G_{Y})$ (Proposition \ref{grammaticaequiv},
folklore) and from two properties of the grammar
$G_Y$, namely the closure of the language $L(G_Y)$ under
concatenation (Lemma \ref{transitivo}) and under
the conjugacy relation (Lemma \ref{closure}),
the latter property being satisfied under the condition
that $Y$ keeps the same property.
Lemma \ref{insertion} is needed in the proof of
Lemma \ref{closure} and the three lemmas show a not
yet thoroughly investigated relation between
insertion, concatenation and closure under
the conjugacy relation.

\begin{lemm} \label{insertion}
For each $w, z \in L(G_{Y})$, for
each $w_1, w_2 \in A^*$ such that
$w = w_1 w_2$, we have
$w_1 z w_2 \in L(G_{Y})$.
\end{lemm}

\begin{propo} \label{grammaticaequiv}
Given a finite set $Y \subseteq A^*$, we have
$Y^{\leftarrow_{*}} = L(G_{Y})$.
\end{propo}

The following two lemmas are needed in the next
section.

\begin{lemm} \label{transitivo}
Given a finite set $Y \subseteq A^*$, the
language $Y^{\leftarrow_{*}}$ is closed under
concatenation, i.e., if $w_1, w_2 \in Y^{\leftarrow_{*}}$
then $w_1w_2 \in Y^{\leftarrow_{*}}$.
\end{lemm}

\begin{lemm} \label{closure}
Let $Y$ be a finite set. If $Y$ is closed
under the conjugacy relation then
$L(G_{Y})$ is closed under the conjugacy
relation.
\end{lemm}

%-----------------------------------

\section{Main Result} \label{MR}

In this section we will state the main results of
the paper. Precisely, in Section \ref{CompleteCSS}
we introduce the notion of a complete system $S$ and we
state that $L$ is generated by $S$ if and only if
$Lin(L)$ may be obtained by iterated insertion starting with
a language closed under the conjugacy relation (Theorem
\ref{MR1}).
A regularity characterization of splicing languages generated
by complete systems follows by the above result (Corollary \ref{MR2}).
Then, in Section \ref{SS} we discuss the case of
$(1,3)$-circular simple systems with only one rule $(a,a)$.
We show that these systems have the same computational power as
complete systems in Proposition \ref{equivalenza}. Consequently
we characterize the corresponding generated languages.

\subsection{Complete circular splicing systems} \label{CompleteCSS}

\begin{de} \label{completeCSSH}
A {\rm complete system} $S = (A, I, R)$ is
a $(1,3)$-CSSH system such that
$R = A \times A$, $alph(I) = A$ and $1 \not \in I$.
\end{de}

\begin{exa} \label{Ex1}
{\rm Let $S = (A, I, R)$, where $A = \{a, b \}$,
$I = \1 \{ab \}$ and $R=\{(a, a), (b, b), (a, b) \}$.
Therefore, $S$ is a complete system. By using
Theorem \ref{MR1}, we will show that $L(S)$ is a
non-regular circular language.
On the contrary, let $S = (A, I', R)$, where
$I' = \1 \{ab, aa, bb \}$. As we will see below, $S$
is a complete system generating a regular circular
language.}
\end{exa}

\begin{rema}
{\rm Let $S = (A, I, R)$ be a complete system.
Notice that $1 \not \in L(S)$ since,
as shown in \cite{rairo}, for any P\u{a}un circular
splicing system $S=(A,I,R)$,
we have that $1 \in I$ if and only
if $1 \in L(S)$.}
\end{rema}

One of our main results, stated in Theorem \ref{MR1},
is based on the following observation. On the
one hand we have already pointed out that if
$L = Y^{\leftarrow_{*}}$ is obtained by iterated insertion,
starting with a finite language $Y$ closed under the
conjugacy relation, then $L$ is closed under concatenation
and under the conjugacy relation (Lemmas \ref{transitivo}, \ref{closure}).
On the other hand, by definition the full linearization
$Lin(L(S))$ of a circular splicing language $L(S)$
is a language closed under the conjugacy relation. In
addition, if $S$ is a complete system then $Lin(L(S))$
is also closed under concatenation (Lemma \ref{transitivo1}).

\begin{lemm} \label{transitivo1}
Let $S$ be a complete system. Then $Lin(L(S))$
is closed under concatenation, i.e., if
$w, w' \in Lin(L(S))$ then $w w' \in Lin(L(S))$.
Analogously, if $\1 w, \1 w' \in L(S)$
then $\1 w w' \in L(S)$.
\end{lemm}

The above result is strengthened by Proposition \ref{P1}.
Indeed we state that $Lin(L(S))$ is the smallest language
closed under conjugacy relation, under concatenation
and containing $I$.

\begin{propo} \label{P1}
Let $S = (A, I, R)$ be a complete system.
Let $L$ be a language such that $L$ is closed under the
concatenation, $L$ is closed under the conjugacy
relation and $Lin(I) \subseteq L$.
Then $Lin(L(S)) \subseteq L$.
\end{propo}

Since the language $L = (Lin(I))^{\leftarrow_{*}}$,
obtained by iterated insertion starting with $Lin(I)$,
satisfies the hypotheses in Proposition \ref{P1},
we obtain Corollary \ref{C1} as a direct result.

\begin{coro} \label{C1}
Let $S = (A, I, R)$ be a complete system.
Then $Lin(L(S)) \subseteq
(Lin(I))^{\leftarrow_{*}} \setminus 1$.
\end{coro}

By the previous preliminary results we can thus prove Theorem
\ref{MR1} stating the connection between pure unitary languages,
pure unitary grammars
and circular splicing languages.

\begin{teo} \label{MR1}
The following conditions are equivalent:

\begin{itemize}
\item[(1)]
There exists a complete system $S = (A, I, R)$ such
that $L = L(S)$.
\item[(2)]
There exists a finite language $Y$, with $alph(Y) = A$,
such that $Y$ is closed under the conjugacy
relation and $Lin(L) = L(G_Y) \setminus 1$,
i.e., $L= \1 (L(G_Y) \setminus 1)$.
\item[(3)]
There exists a finite language $Y$, with $alph(Y) = A$,
such that $Y$ is closed under the
conjugacy relation and
$Lin(L) = Y^{\leftarrow_{*}} \setminus 1$,
i.e., $L= \1 (Y^{\leftarrow_{*}} \setminus 1)$.
\end{itemize}
\end{teo}

\begin{exa} \label{Ex2}
{\rm Let $S = (A, I, R)$ be the complete system
reported in Example \ref{Ex1}, i.e., $A = \{a, b \}$,
$I = \1 \{ab \}$ and $R=\{(a, a), (b, b), (a, b) \}$.
By using Theorem \ref{MR1}, we have $Lin(L(S)) =
\{w \in \{a, b \}^+ ~|~ |w|_a = |w|_b \}$. Thus
$Lin(L(S))$ is a non-regular language and consequently
$L(S)$ is a non-regular circular language.}
\end{exa}

A fundamental consequence of Theorems
\ref{RVP} and \ref{MR1} is stated in
Corollary \ref{MR2}.

\begin{coro} \label{MR2}
Let $S = (A, I, R)$ be a complete system.
Then, $L(S)$ is a regular circular language
if and only if $Lin(I)$ is subword unavoidable.
Consequently, it is decidable whether $L(S)$ is a regular
circular language.
\end{coro}

\begin{exa}
{\rm Let $S = (A, I', R)$ be the complete
system reported in Example
\ref{Ex1}, i.e., $A = \{a, b \}$,
$I' = \1 \{ab, aa, bb \}$ and $R=\{(a, a), (b, b), (a, b) \}$.
Of course $Lin(I')$ is subword unavoidable. Thus,
in view of Corollary \ref{MR2}, $L(S)$ is a regular
circular language. Actually, $Lin(L(S)) = (\{a, b \}^2)^+
= \{w \in \{a, b \}^* ~|~ \exists k > 0: |w| = 2k \}$.}
\end{exa}

\begin{rema}
{\rm Let $S$ be a complete system. Thus
$L = Lin(L(S)) = Lin(L(S))^+$ (Lemma \ref{transitivo1})
and $L \cup \{1 \}$ is a {\it star language}, i.e.,
$L \cup \{1 \}$ is a language which
is closed under the conjugacy
relation and which is the Kleene closure
of a language \cite{rairo}. In \cite{rairo},
the authors defined a class of regular star languages $L$
by means of a property of the finite state automaton recognizing
$L$ and such that $\1 L \in C(Fin,Fin)$.
Finding the relation between this class of star languages
and pure unitary languages is a still unexplored direction.}
\end{rema}

%--------------------------------
\subsection{Simple systems} \label{SS}

In this section we show how our characterization
of the computational power of complete systems,
can be extended to a
special type of $(1,3)$-circular simple systems.
In detail, a reformulation
of Theorem \ref{MR1} and Corollary \ref{MR2}
still hold in terms of $(1,3)$-circular simple systems
having only one rule. This reformulation is obtained
by using a bijection between these systems
and complete systems (Proposition \ref{equivalenza}).
It is clear that, in order to extend this bijection
to $(1,3)$-circular simple systems with a set of rules
of larger size, we need a generalization of the notion
of complete systems.

Let us briefly give an intuitive description of
the results stated in this section.
As already said, the circular
splicing language is obtained by iterated applications
of the splicing operation, starting with all pairs of
circular words in $I$.
Let $S = (A,I,R)$ be a $(1,3)$-circular simple system
where $R = \{(a,a)\}$, $a \in A$.
In this context, given two circular words
$\1 ha$, $\1 ka$, the circular splicing yields
as a result $\1 ha ka$. In other words the splicing
operation is allowed on every position where $a$ appears.
Therefore we handle all words $w$ in $Lin(I)$ (and in
$Lin(\sigma^i(I))$) having the form
$w = x_1 a \cdots x_k a$, where $x_j \in (A \setminus a)^*$.
It is easily seen that each of these words
$w$ is in the free monoid generated by a finite prefix
code $F = \{x_1 a, \ldots , x_k a, \ldots \}$.
Therefore, by a coding morphism $\alpha_j \rightarrow x_j a$,
$F$ is identified with a finite alphabet $A'$ and the
rule $(a,a)$ is identified with the set of rules $A' \times A'$.
Since regular (resp. context-free) languages are closed
under morphism, we can define a bijection between
complete systems $S'$ and $(1,3)$-circular simple systems
$S$ with one rule such that $L(S')$ is regular (resp.
context-free) if and only if $L(S)$ is regular (resp.
context-free). Let us state the definitions and results
in a precise way.

Given
$I = \1 \{w_{1}, \ldots , w_{k} ~|~ 1 \leq i \leq k\}$,
the set $F \subseteq A^+$ associated
with $I$ is the set of smallest cardinality
which satisfies the following conditions:
(1) for each word $w$ in $F$ we have $w \in A^*a$ and $|w|_a =1$;
(2) for each $\1 w \in I$, if $w'$ is any linearization
of $\1 w$ such that $w' \in A^*a$ then $w' \in F^+$.

Hence $F = \{x_ia ~|~ 1 \leq i \leq n \}$ is
a finite prefix code (i.e., $F \cap FA^+ = \emptyset$)
and for each $\1 w_i \in I$, we have
$\1 w_i = \1 x_{i,1}a \cdots x_{i,j_i}a$,
where $j_i \in \mathbb{N}$, $|x_{i,g}|_a = 0$, for each
$1 \leq g \leq j_i$ and $x_{i,1}a, \ldots, x_{i,j_i}a$
is a set of not necessarily different elements in $F$.
Here, we assume that $a \in alph(w)$, for each
$w \in I$. Notice that, in order to characterize the circular
splicing language generated by $S$, there is no loss of generality
in making this assumption. Indeed, by
using a result stated in \cite{DFFZtcs}, we have that
$L(S) = L(S_1) \cup (I \setminus I_1)$, where
$S_1 = (A, I_1, R)$ and $I_1 = \{ w \in I ~|~ a \in alph(w) \}$.
From now on, $S = (A,I,R)$ will denote a
$(1,3)$-circular simple system with
$R = \{(a,a)\}$, $a \in A$ and
$I = \1 \{x_{i,1}a \cdots x_{i,j_i}a ~|~ 1 \leq i \leq k\}$,
where $|x_{i,g}|_a = 0$, for each $1 \leq g \leq j_i$.
Of course $L(S) \subseteq \1 F^+$.
We now state that the above systems
have the same computational power as special
complete systems.

Let $\varphi$ be a coding morphism for $F$,
i.e., a morphism $\varphi \colon A'^* \rightarrow A^*$
which is injective and such that $F = \varphi(A')$
\cite{bpr}. Thus, $|A'| = |F|$. Set
$A' =\{\alpha_1, \ldots, \alpha_n \}$. We know
that $\varphi$ is defined by
$\varphi(\alpha_i) = x_i a$, $1 \leq i \leq n$.
We also know that $\varphi$ defines a map
from $\1 A'^*$ into $\1 A^*$ if we
set $\varphi(\1 w)=\1 \varphi(w)$, for all
$w \in A'^*$, i.e., $\varphi(\1 w)$ does not
depend on which representative in $\1 w$ we
choose to define it by \cite{hb}. Thus,
if $w,w' \in A'^*$ and
$w \sim w'$ then $\varphi(w) \sim \varphi(w')$.
In what follows $S' = (A', I', R')$ will
denote the complete system defined
by $I' = \varphi^{-1}(I)$. Thus, $A' = alph(I')$
and $R' = A' \times A'$.

\begin{propo} \label{equivalenza}
We have $\varphi(L(S')) = L(S)$
and $\varphi^{-1}(L(S)) = L(S')$.
Consequently, $L(S)$ is a regular (resp. context-free)
circular language if and only if $L(S')$ is
regular (resp. context-free).
\end{propo}

In view of Proposition \ref{equivalenza}, Theorem
\ref{MR1} and Corollary \ref{MR2} may be
rephrased for $(1,3)$-circular simple systems $S$
with only one rule (see \cite{estesa}).

\begin{exa} \label{ultimo}
{\rm We recall that a word $x \in A^+$
is called unbordered if $x \in u A^+ \cap A^+ u$
implies $u = 1$ \cite{bpr}.
Given $w,x \in A^*$, with $x$ being an
unbordered word, we denote by $|w|_x$
the number of occurrences of $x$ in $w$.
Let $S = (A,I,R)$ be the $(1,3)$-circular simple system
defined by $A = \{a, b, c\}$,
$I= \1 \{baca\}$ and $R = \{(a,a) \}$.
In view of Proposition \ref{equivalenza} we
have $L(S) = \varphi(L(S'))$, where
$S' = (A', I', R')$ is the complete system defined
by $I' = \1 \{ \alpha \beta \}$, $A' = alph(I') =
\{ \alpha, \beta \}$ and $R' = A' \times A'$.
By using Theorem \ref{MR1}, we have 
$Lin(L(S')) = 
\{w \in \{\alpha, \beta \}^+ ~|~ |w|_{\alpha} = |w|_{\beta} \}$
and so $L(S') =
\1 \{w \in \{\alpha, \beta \}^+ ~|~ |w|_{\alpha} = |w|_{\beta} \}$
is a (non-regular) context-free circular
language (see Example \ref{Ex2}).
By using Proposition \ref{equivalenza} 
once again, $L(S) = \varphi(L(S')) = 
\1 \{w \in \{ a,b,c \}^+ ~|~ |w|_{ba} = |w|_{ca} \}$
is a (non-regular) context-free circular
language. An ad hoc proof of the non-regularity
of the circular language $L(S)$ has also been
reported in \cite{survey}.}
\end{exa}

\begin{center}
{\bf ACKNOWLEDGEMENTS}
\end{center}
The authors are very grateful to an
anonymous referee for
pointing out reference \cite{EHRtcs}.

%-------------------------------

\end{document}